\documentclass[10pt,journal,twocolumn,twoside]{IEEEtran} 
\newcommand{\figref}[1]{Fig.~\ref{#1}}
\usepackage{titlesec}
\usepackage{stfloats}	
\usepackage{graphicx}
\usepackage{epstopdf}
\usepackage{float}
\usepackage{cite}
\usepackage{graphicx}
\usepackage{subfigure}
\usepackage{algorithm,algorithmic}
\usepackage{array}
\usepackage{amsmath}
\usepackage{amssymb}
\usepackage{mdwmath}
\usepackage{CJK}
\usepackage{mdwtab}
\usepackage{eqparbox}
\usepackage{fixltx2e}
\usepackage{cases}
\usepackage{bm}
\usepackage{multirow}
\usepackage{psfrag}
\usepackage[usenames]{color}
\usepackage{setspace}
\usepackage{fixmath}
\usepackage{mathdots}
\usepackage{amsthm}  
\usepackage{graphicx}
\usepackage{subfigure} 
\usepackage{overpic}
\usepackage{amssymb}

\usepackage{xcolor} 

\usepackage{color}
\usepackage{makecell}
\begin{document}
\title{ Social-Mobility-Aware Joint Communication and Computation Resource Management in NOMA-Enabled Vehicular Networks}
	\author{
	Tong~Xue,
	Haixia~Zhang,~\IEEEmembership{Senior~Member,~IEEE,} Hui~Ding, and 
	Dongfeng Yuan,~\IEEEmembership{Senior~Member,~IEEE}
	\thanks{T. Xue, H. Zhang, H. Ding and D. Yuan are all with Shandong Key Laboratory of Wireless Communication Technologies, Shandong University, Jinan, Shandong, 250061, China. 
		T. Xue and H. Zhang are also with School of Control Science and Engineering, Shandong University, Jinan, Shandong, 250061, China (e-mail: tong\_xue@mail.sdu.edu.cn; haixia.zhang@sdu.edu.cn).}
}

\maketitle

\begin{abstract}
	The existing computation and communication (2C) optimization schemes for vehicular edge computing (VEC) networks mainly focus on the physical domain without considering the influence from the social domain. This may greatly limit the potential of task offloading, making it difficult to fully boom the task offloading rate with given power, resulting in low energy efficiency (EE). To address the issue, this letter devotes itself to investigate social-mobility-aware VEC framework and proposes a novel EE-oriented 2C assignment scheme. In doing so, we assume that the task vehicular user (T-VU) can offload computation tasks to the service vehicular user (S-VU) and the road side unit (RSU) by non-orthogonal multiple access (NOMA). An optimization problem is formulated to jointly assign the 2C resources to maximize the system EE, which turns out to be a mixed integer non-convex objective function. To solve the problem, we transform it into separated computation and communication resource allocation subproblems. Dealing with the first subproblem, we propose a social-mobility-aware edge server selection and task splitting algorithm (SM-SSTSA) to achieve edge server selection and task splitting. Then, by solving the second subproblem, the power allocation and spectrum assignment solutions are obtained utilizing a tightening lower bound method and a Kuhn-Munkres algorithm. Finally, we solve the original problem through an iterative method. Simulation results demonstrate the superior EE performance of the proposed scheme. 
\end{abstract}

\maketitle

\section{Introduction}
\IEEEPARstart{W}{ith} the booming development of intelligent vehicles and wireless communications, a variety of advanced vehicular entertainment services such as high-definition map have emerged in vehicular networks. Quite a lot emerging vehicular entertainment services are computationally-intensive, but the vehicular users (VUs) with constrained computation capability can not satisfy the quality of service (QoS) of such services. To overcome this, it is paramount crucial to utilize vehicular edge computing (VEC) technology that leverages the abundant computation resources at proximity edge servers (i.e., road side units (RSUs) and idle service vehicular users (S-VUs)) \cite{9978912}. However, when the VUs offload tasks to the edge servers, the power consumption increases significantly. Improving the transmission rate of offloaded tasks with limited power, i.e. energy efficiency (EE), has become a major concern in VEC networks. One feasible method is to optimize the communication resources, such as spectrum and power. In addition, designing appropriate task computation policies, such as determining where to offload the computational tasks, is another way to enhance the EE \cite{9638386}. There are works focusing on joint optimizing communication and computation (2C) resource allocation strategies to maximize the EE in orthogonal multiple access (OMA)-enabled VEC networks \cite{9698985,9709889,9560158}. 

In addition, non-orthogonal multiple access (NOMA) has also been regarded as a potential technology to further enhance the system EE \cite{8636993}. With the help of successive interference cancellation (SIC) at the receiver, co-channel interference can be suppressed, which enhances the system sum-rate and finally achieves the significant improvement of the system EE. Therefore, there are works focusing on optimizing 2C resources by integrating NOMA into VEC networks\cite{8600310,9470949,9625230}. For instance, Cheng \textit{et al.} \cite{9470949} proposed a joint optimization strategy for binary task splitting and power control to maximize EE, where the task VU (T-VU) can offload its computation task to the S-VU or RSU by NOMA. With the same goal, based on the minimum distance S-VU selection (MDSS) strategy, Wen \textit{et al.} studied a NOMA-enabled three-sided matching theory to jointly optimize the task splitting and power control in cognitive vehicular networks \cite{9625230}. Literature \cite{8600310,9470949,9625230} focused on the 2C optimization strategies based on the physical domain without considering the influence from the social domain. This may greatly limit the potential of task offloading, making it difficult to fully boom the task offloading rate with given power, resulting in a low EE. Therefore, it is indispensable to improve the system EE by designing a  social-mobility-aware 2C optimization strategy.

Inspired by the aforementioned analysis, this work designs a social-mobility-aware VEC framework and proposes a novel EE-oriented 2C assignment scheme. In doing so, we assume the T-VU offloading computation tasks to S-VU and RSU by NOMA. Meanwhile, to improve the resource utilization, we enable T-VUs to reuse the spectrum resource with cellular users (CUs). An optimization problem is formulated to jointly allocate the 2C resource to maximize the system EE, while guaranteeing the QoS requirements of all CUs and T-VUs. The formulated optimization problem is a mixed integer non-convex. To solve this problem, we decompose it into separated computation and communication resource allocation subproblems. To deal with the computation subproblem, we propose a social-mobility-aware edge server selection and task splitting algorithm (SM-SSTSA) to determine edge server and task splitting. Then, by solving the communication subproblem, the power allocation and spectrum assignment solutions are obtained by using a tightening lower bound method and a Kuhn-Munkres algorithm. Finally, we solve the original problem by iteratively solve the two subproblems. Simulation results demonstrate the superiority of the proposed scheme in terms of the EE.

\section{System Model and Problem Formulation}
\begin{figure}[t]
	\centering
	\includegraphics[width=0.45\textwidth,height=0.22\textwidth]{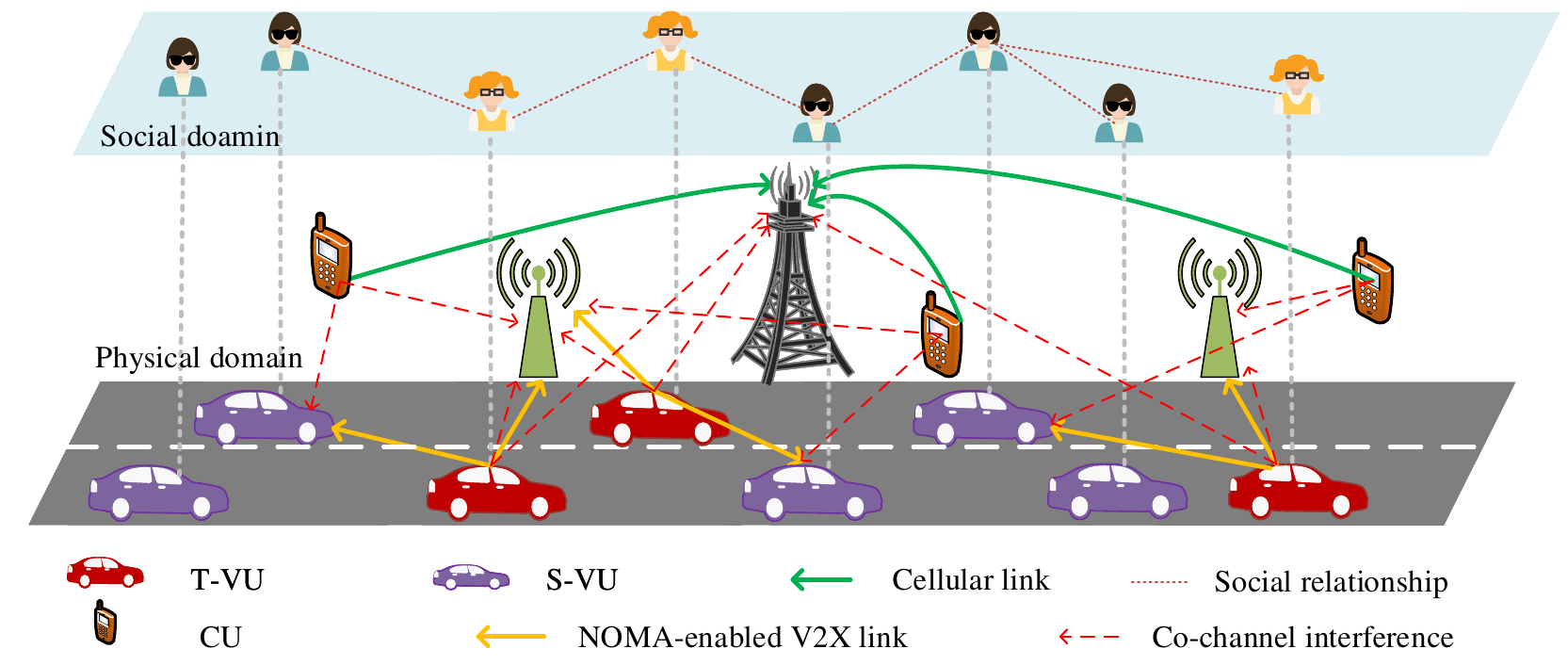}\\
	\caption{System model for NOMA-enabled social-mobility-aware VEC networks. }
	\label{SystemModel}
\end{figure}
\subsection{Physical and Social Domain Model }
This work studies a social-mobility-aware VEC network that utilizes NOMA technology to ensure the differentiated QoS requirements for each CU and T-VU, as shown in \figref{SystemModel}. In the physical domain, a macro base station (MBS) is deployed to support high-rate data transmission of $U$ CU indexed by $u\in\mathcal{U}=\{1,2,...,U\}$, and $S$ RSUs indexed by $s\in\mathcal{S}=\{1,2,...,S\}$ with coverage radius $r$ are deployed to support the computationally-intensive services of $M$ T-VU indexed by $m\in\mathcal{M}=\{1,2,...,M\}$. Each RSU is equipped with a mobile edge computing (MEC) server. Given the limited computation capability of T-VUs, we allow the T-VUs to offload computational tasks to the proximity RSUs through vehicle-to-infrastructure (V2I) links and the idle S-VUs through vehicle-to-vehicle (V2V) links. It is assumed that there are $N$ idle S-VUs indexed by $n\in\mathcal{N}=\{1,2,...,N\}$. Based on the characteristic of task offloading, this work allows the T-VU offloading tasks to the RSU server and the S-VU by utilizing NOMA. In the social domain, leveraging social relationships can help build trustworthy V2V offloading links and improve the effective task offloading rate with limited power, i.e., EE \cite{Feng-8316777}. In this work, the social relationship graph among VUs is denoted by $\mathcal{G}=(\bm{Z},\bm{\delta})$, where $\bm{Z}$ denotes the set of all VUs with $\bm{Z}=\mathcal{M}\cup\mathcal{N}$, and $\delta_{m,n}\in\bm{\delta}=\big\{\delta_{1,1},\delta_{1,2},...\delta_{M,N}\big\}$ is a binary variable representing the social relationship between the $m$th T-VU and the $n$th S-VU. If the $m$th T-VU agrees to share computation task with the $n$th S-VU, then $\delta_{m,n}=1$, otherwise, $\delta_{m,n}=0$. 

\subsection{Communication and Computation Model}
 In the NOMA-enabled VEC network, it is assumed that there are totally $F$ available sub-channels (SCs) indexed by $f\in\mathcal{F}=\{1,2,…,F\}$. Without loss of generality, we assume $F = U$, and each CU uses a single SC. To improve the spectrum resource utilization, the CUs and the T-VUs are allowed to share the spectrum band. It is assumed that only one V2I link and one V2V link utilize NOMA mode to share the SC occupied by one CU. Therefore, the signal-to-interference-plus-noise ratio (SINR) of the $u$th CU at the time slot $t$, $t\in\mathcal{T}=\{1,2,...,T\}$, can be expressed as
\begin{equation}
R_{u}(t)=\sum\limits_{f\in\mathcal{F}}B\log_2(1+ \frac{ P_{u}^{op}(t)X_{u,f}(t)H_{u}(t)}
{\sum\limits_{m\in\mathcal{M}}Q_{1}+\sigma^2}),
\end{equation}
where $B$ represents the bandwidth of each SC, $Q_{1}={( \epsilon_{m,1}(t)+\epsilon_{m,2}(t))P_{m}^{th}X_{m,f}(t)H_{m,u}(t)}$, with $\epsilon_{m,1}(t)$ and $\epsilon_{m,2}(t)$ represent the power allocation coefficients from the $m$th T-VU to the RSU and to the S-VU at the $t$th time slot, respectively, $P_m^{th}$ is the maximum transmit power of the $m$th T-VU, $P_u^{op}(t)$ denotes the optimal transmit power of the $u$th CU at the $t$th time slot, $\sigma^2$ is the noise power, the binary variable $X_{m,f}(t)\in\{0,1\}$ is defined as the spectrum assignment factor. If the $m$th T-VU occupies the $f$th SC at the $t$th time slot, $X_{m,f}(t)=1$, otherwise, $X_{m,f}(t)=0$. Similarly, $X_{u,f}(t)$ is also a spectrum assignment indicator of the $u$th CU at the $t$th time slot. $H_u(t)$ and $H_{m,u}(t)$ are the channel and interference channel power gain of the $u$th CU at the $t$th time slot.

For each NOMA-enabled V2V link and V2I link's receiver, it assumes that each receiver is able to decode the received messages via SIC, and the decoding order is based on the increasing order of channel coefficients. If $H_{m,s}<H_{m,n}$, the $m$th T-VU tends to allocate higher power to the $s$th RSU than that of the $n$th S-VU, such that $\epsilon_{m,1}>\epsilon_{m,2}$. Through the NOMA protocol, the $m$th V2I receiver is firstly decoded. The $m$th V2V link is then decoded and the co-channel interference from the $m$th V2I link is removed\footnote{If $H_{m,s}>H_{m,n}$, the $m$th V2V link will be firstly decoded, and the SINR of receiver will be changed.} by SIC. Therefore, the SINR of the $m$th V2I link's receiver (i.e, the $s$th RSU) at the $t$th time slot can be expressed as
	\begin{equation}
	\label{gamma_m,1,f}
R_{m,s}(t)=\sum\limits_{f\in\mathcal{F}}B\log_2(1+
\underbrace{ \frac{ \epsilon_{m,1}(t)P_{m}^{th}X_{m,f}( t)H_{m,s}(t)}
	{Q_2+\sigma^2} }_{\gamma_{m,f}(t)} ),
	\end{equation}
where $Q_2={\sum_{u\in\mathcal{U}}P_{u}^{op}(t)X_{u,f}(t)H_{u,s}(t)}+\epsilon_{m,2}(t)P_{m}^{th}X_{m,f}(t)H_{m,s}(t)$. The SINR of the $m$th V2V link's receiver (i.e., the $n$th S-VU) at the $t$th time slot can be expressed as
	\begin{equation}
\label{gamma_m,n,2,f}
R_{m,n}(t)=\sum\limits_{f\in\mathcal{F}}B\log_2(1+ \underbrace{\frac{ X_{m,f}(t)\Psi_{m,n}{(t)}Q_3}
	{Q_4+\sigma^2} }_{\gamma_{m,n,f}(t)}),
\end{equation}
where $Q_3=\epsilon_{m,2}(t)P_{m}^{th}H_{m,n}(t)$, $Q_4=\sum_{u\in\mathcal{U}}{P_{u}^{op}(t)X_{u,f}(t)\Psi_{m,n}{(t)}H_{u,n}(t)}$, $H_{m,s}(t)$ and $H_{m,n}(t)$ are the channel power gains from the $m$th T-VU to the $s$th RSU server and to the $n$th S-VU at the $t$th time slot, respectively, $H_{u,s}(t)$ denotes the interference channel power gain from the $u$th CU to the $s$th RSU at the $t$th time slot, $H_{u,n}(t)$ is the interference channel power gain from the $u$th CU to the $n$th S-VU at the $t$th time slot. The binary variable $\Psi_{m,n}(t)$ composed of both mobility and social relationships is denoted as
	\begin{equation}
\Psi_{m,n}(t)=k_{m, n}(t)\cdot \delta_{m,n}(t),
\end{equation}
where $k_{m, n}(t)$ is the mobility relationship between the $m$th T-VU and the $n$th S-VU at the $t$th time slot, where
	\begin{equation}
\label{k_mn}
\begin{aligned}
k_{m, n}(t)=
\left\{\begin{array}{l}
1,
\text { if } \rho_{m, n}(t)<\zeta_{th},\\
0, \text { otherwise},
\end{array}\right.     
\end{aligned}
\end{equation}where $\zeta_{th}$ represents the threshold of physical domain, and $\rho_{m,n}(t)$ is written as
\begin{equation}
\begin{aligned}
\rho_{m,n}(t)=&\psi \cdot f(\Delta d_{m,n}(t))+(1-\psi) \cdot f(\Delta v_{m,n}(t)),
\end{aligned}
\end{equation}where $\Delta d_{m,n}(t)$ is the distance between T-VU and S-VU, $\Delta v_{m,n}(t)$ represents the difference in velocity between T-VU and S-VU, $\psi\in[0,1]$ is the weight of the distance,  $f(\cdot)$ is the normalized function.

	\begin{algorithm}[t]     
	\caption{Social-Mobility-Aware Edge Server Selection and Task Splitting Algorithm (SM-SSTSA)}
	\label{alg:PR1}
	\begin{algorithmic}[1]
		\STATE Initialize the set of the social-ties $\boldsymbol\delta$. 
		\FOR {$m=1:M$}
		\STATE Calculate $k_{m,n}(t)$ based on (\ref{k_mn}).
		\STATE Find the S-VU $n',n'\in\mathcal{N}$ that makes $\delta_{m,n'}(t)\cdot k_{m,n'}(t)=1$, and place it into the alternative S-VUs set $\mathcal{A}_m(t)$.
		\STATE Calculate $R_{total}(t)-\xi P_{total}(t)$, sort $n'$ in a decreasing order of $\varphi_{m,n'}(t)$, and update the order of $n'$ in $\mathcal{A}_m(t)$.
		\FOR {$i=\mathcal{A}_m(t)$}
		\STATE Substitute $i$ into (\ref{Con_delay}).
		\IF {it exits $\beta_{m,1}(t)^*,\beta_{m,2}(t)^*$ satisfying (\ref{Con_delay})} 
		\STATE Break.
		\ENDIF	
		\ENDFOR
		\STATE $n^*=i$, $\Psi_{m,n^*}{(t)}=1$.
		\STATE Delete $n^*$ from the set $\mathcal{N}$. 
		\ENDFOR
	\end{algorithmic}
\end{algorithm}
We define a tuple $(D_m(t), C_m,\boldsymbol{\beta_m}(t))$ to characterize the task of the $m$th T-VU at the $t$th time slot, where $D_m(t)$ is the size of the computation task, $C_m$ is the number of CPU cycles required for computing 1-bit data, $\boldsymbol{\beta_m}(t)=\{\beta_{m,1}(t),\beta_{m,2}(t)\}\in[0,1]$, $\beta_{m,1}(t)$ represents the computing task splitting factor from the $m$th T-VU to the RSU server. $\beta_{m,2}(t)$ is the computed ratio by the S-VU. Thus, $(1-\beta_{m,1}(t)-\beta_{m,2}(t))$ denotes the portion of the computing task left for local executing (i.e., the $m$th T-VU). Therefore, the task executing delay at the $m$th T-VU is
\begin{equation}
\label{Con_tvu}
\frac{D_m(t)(1-\beta_{m,1}(t)-\beta_{m,2}(t)) C_m}{y_{m}}<T_{tol},
\end{equation}
where $y_{m}$ (in CPU cycle/s) is the assigned computing resource for executing local tasks, $T_{tol}$ denotes the maximum tolerant delay of each T-VU. 

The task offloading and executing delay from the $m$th T-VU to the RSU server and to the $n$th S-VU can be expressed as
\begin{equation}
\label{Con_rate_v2i}
\frac{D_m{(t)}\beta_{m,1}(t)}{R_{m,s}(t)}+\frac{D_m(t)\beta_{m,1}(t) C_m}{y_{m,s}}<T_{tol},
\end{equation}
\begin{equation}
\label{Con_rate_v2v}
\frac{D_m(t)\beta_{m,2}(t)}{R_{m,n}(t)}+\frac{D_m(t)\beta_{m,2}(t)C_m}{y_{m,n}}<T_{tol},
\end{equation}
where $y_{m,s}$ (in CPU cycle/s) and $y_{m,n}$ (in CPU cycle/s) are the computing resource allocated to the $m$th T-VU served by the RSU server and the $n$th S-VU, respectively. The EE of the NOMA-enabled VEC networks is expressed as
\begin{equation}
\label{EE}
\xi=\sum\limits_{t\in\mathcal{T}}\frac{R_{total}(t)}{P_{total}(t)}=\sum\limits_{t\in\mathcal{T}}\sum\limits_{n\in\mathcal{N}}\sum\limits_{m\in\mathcal{M}}\sum\limits_{s\in\mathcal{S}}\frac{R_{m,s}(t)+R_{m,n}(t)}{P_{cir}+\tilde{P}_{m,n,s}(t)},
\end{equation}
where $\tilde{P}_{m,n,s}(t)=\kappa y_m^3+\epsilon_{m,1}(t)P_m^{th}+\kappa y_{m,s}^3+\epsilon_{m,1}(t)P_m^{th}+\kappa y_{m,n}^3$, $\kappa$ is the effective switched capacitance depending on the CPU architecture, and $P_{cir}$ is the circuit power consumption.

\subsection{Problem Formulation}
In this work, our objective is to maximize the EE for task offloading of the NOMA-enabled VEC network by optimizing the edge server selection $\boldsymbol{\Psi}$, the task splitting $\boldsymbol{\beta}$, the spectrum assignment $\boldsymbol{X}$ and the power allocation $\boldsymbol{\epsilon}$. Notably, $\boldsymbol{\Psi}$, $\boldsymbol{\beta}$, $\boldsymbol{X}$ and $\boldsymbol{\epsilon}$ are matrices composed of variables $\Psi_{m,n}(t)$, $\{\beta_{m,1}(t),\beta_{m,2}(t)\}$, $\{X_{u,f}(t), X_{m,f}(t)\}$ and $\{\epsilon_{m,1}(t),\epsilon_{m,2}(t)\}$, respectively. Mathematically, the problem is formulated as
\begin{subequations}
	\label{P1}
	\begin{align}
	&\mathcal{P}1:\max\limits_{\substack{\Psi_{m,n}(t),\beta_{m,1}(t),\beta_{m,2}(t),X_{u,f}(t),\\ X_{m,f}(t),\epsilon_{m,1}(t),\epsilon_{m,2}(t)}} \xi        \label{obj}\\	
	\textrm{s.~t.~} 
    &\Psi_{m,n}{(t)}, X_{m,f}{(t)}, X_{u,f}{(t)}\in\{0,1\},\forall m,s,n,u,f,t,\label{Con_binariy1}\\
    &0\leq\beta_m(t)\leq1,\forall m,t, \label{Con_binariy2}\\
    	& \epsilon_{m,1}(t)\geq 0, \epsilon_{m,2}(t)\geq 0,\epsilon_{m,1}(t)+\epsilon_{m,2}(t)\leq 1,\forall m,t,\label{Con_variable1}\\
	&	R_{u}(t)\geq R_{th,u},\forall f,u,t,\label{Con_rate_cellular}\\
		& \sum\limits_{f\in\mathcal{F}}X_{u,f}{(t)}= \sum\limits_{f\in\mathcal{F}}X_{m,f}{(t)}=1, \forall u,m,t,\label{Con_RB1}\\
	& \sum\limits_{u\in\mathcal{U}}X_{u,f}{(t)}=1, \sum\limits_{m\in\mathcal{M}}X_{m,f}{(t)}\leq1,\forall f,t,\label{Con_RB2}\\
			& (\ref{Con_tvu}),(\ref{Con_rate_v2i}),(\ref{Con_rate_v2v}),\label{Con_delay}
	\end{align}
\end{subequations}
where $R_{th,u}$ is the minimum data rate thresholds for the $u$th CU, constraints (\ref{Con_binariy2})-(\ref{Con_variable1}) list the feasible task splitting and power allocation of the T-VUs, respectively, constraint (\ref{Con_rate_cellular}) represents the QoS requirements of CUs, constraint (\ref{Con_RB1}) restricts that each user (T-VU and CU) can only access to one SC, each SC can be shared by one CU and at most one T-VU according to constraint (\ref{Con_RB2}).

It is obvious that (\ref{obj}) is a fractional programming, which can be converted into a subtractive form \cite{8695055}. Therefore, (\ref{obj}) is reformulated as
\begin{equation}\label{transformation_P}
\max\limits_{\substack{\Psi_{m,n}(t),\beta_{m,1}(t),\beta_{m,2}(t),X_{u,f}(t),\\ X_{m,f}(t),\epsilon_{m,1}(t),\epsilon_{m,2}(t)}} 
\sum\limits_{t\in\mathcal{T}}(R_{total}(t)-\xi P_{total}(t)).
\end{equation}
\section{Solution of the EE Optimization Problem}
Since the communication and computation resource decision of $\mathcal{P}1$ is made in each time slot and there is no interdependence among time slots, we transform the optimization problem across the whole time slots into one time slot optimization problem. But, the obtained one time slot optimization problem is still non-convex, and it is difficult to obtain the global optimal solution. As an alternative, we decompose it into 1) computation resource optimization subproblem $\mathcal{P}2$ and 2) communication resource optimization subproblem $\mathcal{P}3$. $\mathcal{P}2$ and $\mathcal{P}3$ can be given by
\begin{subequations}
	\label{P3}
	\begin{align}
	\mathcal{P}2:&~\max\limits_{\Psi_{m,n}(t),\beta_{m,1}(t),\beta_{m,2}(t)}
	(R_{total}(t)-\xi P_{total}(t))      \label{P3_obj}\\
	\textrm{s.~t.~} 
	&~ \Psi_{m,n}{(t)},\Psi_{m,s}{(t)}\in\{0,1\},\forall m,s,n,\label{P3_b}\\
	&~ (\ref{Con_binariy2}), (\ref{Con_delay}),
	\end{align}
\end{subequations}
\begin{subequations} 
	\label{p2.2}
	\begin{align}
	\mathcal{P}3:&\max\limits_{X_{u,f}(t),X_{m,f}(t),\epsilon_{m,1}(t),\epsilon_{m,2}(t)}(R_{total}(t)-\xi P_{total}(t))            \label{p3_obj}\\	
	\textrm{s.~t.~} &~X_{m,f}{(t)}, X_{u,f}{(t)}\in\{0,1\},\forall m,u,f, \\
	&~ (\ref{Con_rate_v2i}), (\ref{Con_rate_v2v}), (\ref{Con_variable1})-(\ref{Con_RB2}).\label{p4_Con}
	\end{align}
\end{subequations}
It is seen that $\mathcal{P}2$ is NP-hard. To find a tractable solution, we design a heuristic SM-SSTSA as shown in \textbf{Algorithm 1}. Then, to solve the communication resource allocation subproblem, we decouple $\mathcal{P}3$ into a power allocation subproblem and a spectrum assignment subproblem, which can be solved iteratively.

As proved in \cite{8014491}, when the spectrum assignment variable is fixed, (\ref{p3_obj}) is written as 
	\begin{equation}	    
\begin{aligned}
\label{obj_rewritten1}
&R_{m,s}(t)+R_{m,n}(t)-\xi P_{total}(t)\\
\geq&   \underbrace{ b_1\log_2\gamma_{m,f}(t)+c_1 }_{\Phi_{1}(t)}   +   \underbrace{b_2\log_2\gamma_{m,n,f}(t)+c_2 }_{\Phi_{2}(t)}    -\xi P_{total}(t),
\end{aligned}
\end{equation}
where $b_1$, $b_2$, $c_1$ and $c_2$ are
\begin{equation}
\label{b_1}
b_1=\frac{\tilde{\gamma}_{m,f}(t)}{1+\tilde{\gamma}_{m,f}(t)}, b_2=\frac{\tilde{\gamma}_{m,n,f}(t)}{1+\tilde{\gamma}_{m,n,f}(t)},
\end{equation}
\begin{equation}
\label{c_1}
c_1=\log_2(1+\tilde{\gamma}_{m,f}(t))-\frac{\tilde{\gamma}_{m,f}(t)}{1+\tilde{\gamma}_{m,f}(t)}\log_2 \tilde{\gamma}_{m,f}(t),
\end{equation}
\begin{equation}
\label{c_2}
c_2=\log_2(1+\tilde{\gamma}_{m,n,f}(t))-\frac{\tilde{\gamma}_{m,n,f}(t)}{1+\tilde{\gamma}_{m,n,f}(t)}\log_2 \tilde{\gamma}_{m,n,f}(t).
\end{equation}

Then, the lower bound of the objective function in (\ref{p3_obj}) can be written as 
\begin{equation}
\max\limits_{\boldsymbol{\epsilon}}  ( {\Phi_{1}}(t)   +  {\Phi_{2}}(t)    -\xi P_{total}(t)).
\end{equation}

Denote $\epsilon_{m,1}(t)=2^{w_{m,1}(t)}$ and $\epsilon_{m,2}(t)=2^{w_{m,2}(t)}$, the power control subproblem can be rewritten as
\begin{subequations}
	\label{P8}
	\begin{align}
	\mathcal{P}4:&~\max\limits_{w_{m,1}(t),w_{m,2}(t)}  \Big( {\Phi_{1}}(t)   +  {\Phi_{2}}(t)    -\xi P_{total}(t)\Big) \label{P3a}\\
	\textrm{s.~t.~} 
	&~ 2^{w_{m,1}(t)}\geq 0, 2^{w_{m,2}(t)}\geq 0,\forall m,\label{}\\
	&~2^{w_{m,1}(t)}+2^{w_{m,2}(t)}\leq 1,\forall m,\label{P7_Con_w} \\
	&~ (\ref{Con_rate_v2i}),(\ref{Con_rate_v2v}),(\ref{Con_rate_cellular}).\label{} 
	\end{align}
\end{subequations}	

Since (\ref{P8}) is a standard convex optimization problem, we adopt Lagrange dual decomposition to solve it. 

Given $\boldsymbol{\epsilon}$, the spectrum assignment subproblem is a complicated matching among CUs, T-VUs and SCs, which is proved to be NP-hard. From (\ref{Con_RB1})-(\ref{Con_RB2}),  the relationship between the cellular user and SC belongs to one-to-one match. To facilitate the solution, the complex match among CUs, T-VUs and SCs is transformed into the new match among CUs and T-VUs. The new spectrum assignment variable between the $u$th CU and the $m$th T-VU at the $t$th time slot is denoted as $X_{u,m}(t)$. Therefore, the spectrum assignment subproblem can be rewritten as
\begin{subequations} 
	\label{p4}
	\begin{align}
	\mathcal{P}5:&~\max\limits_{X_{u,m}{(t)}} \Big(R_{total}(t)-\xi P_{total}(t)\Big)     \label{p4_obj_RB}\\
	\textrm{s.~t.~}
	&X_{u,m}{(t)}\in \{0,1\},\forall  u,m,\\
	& \sum\limits_{m\in\mathcal{M}}X_{u,m}{(t)}\leq1, \forall u,\\
	&\sum\limits_{u\in\mathcal{U}}X_{u,m}{(t)}=1,\forall m,
	\end{align}
\end{subequations}
which can be solved by a Kuhn-Munkres algorithm.

To solve $\mathcal{P}1$, JCCRAA is proposed as shown in \textbf{Algorithm 2}, which composes of solving the computation resource allocation subproblem and the communication resource allocation subproblem. In \textbf{Algorithm 2}, by solving \textbf{Algorithm 1}, $\boldsymbol{\Psi}$ and $\boldsymbol{\beta}$ can be obtained. Then,  the analytical expression of $\boldsymbol{X}$ and $\boldsymbol{\epsilon}$ can be derived by using the tightening lower bound method  and the Kuhn-Munkres algorithm. Next, by substituting the obtained assigned spectrum and allocated power into \textbf{Algorithm 1}, the S-VUs selection and task splitting strategies are updated. Repeat the process until convergence, the original problem is solved.
\begin{algorithm}[t]     
	\caption{Joint Communication and Computation Resource Allocation Algorithm (JCCRAA)}
	\label{alg:PR3}
	\begin{algorithmic}[1]
		\FOR {$t=1:T$}
		\STATE Initialize the index $i=0$, the power splitting $w_{m,1}^{\{i\}}(t)$, $w_{m,2}^{\{i\}}(t)$, the maximum number of iterations $L_1$ and the convergence threshold $\varepsilon$. $\xi^{(0)}=0$.
		\WHILE {$\alpha^{(i)}\geq\varepsilon$ and $i<L_1$}
		\STATE  $i=i+1$.
		\STATE Calculate $\boldsymbol{\Psi}$ and $\boldsymbol{\beta}$ based on \textbf{Algorithm 1}.
		\STATE Solve $\boldsymbol{X}$ and $\boldsymbol{\epsilon}$ by using the tightening lower bound method and the Kuhn-Munkres algorithm.		
		\STATE Update $\xi^{(i)}$ via (\ref{EE}).
		\STATE $\alpha^{(i)}=|\xi^{(i)}-\xi^{(i-1)}|$.
		\ENDWHILE
		\ENDFOR
	\end{algorithmic}
\end{algorithm}

\section{Simulation Results and Analysis}
Intensive simulations are done to show the performance of the proposed algorithm.  It is assumed that all the users are located within a target rectangular area 1000 m$\times$ 1000 m. The simulation parameters are set according to 3GPP TR 36.885 \cite{3GPP}, where a MBS is located at the center of the area and a number of RSUs with $r= 150$ m are located at the roadside in the area. The number of lanes is	6, and the width of each lane is 4 m. The average inter-VU distance driving in the same lane is $2.5v$ m with $v$ representing the moving speed of vehicles in meter per second. Besides, we set $P_u^{op}=20$ dBm, $P_m^{th}= [15, 30]$ dBm, $D_m=[10^4,10^5]$ bits.  

\begin{figure}[t]
	\centering
	\includegraphics[width=0.45\textwidth,height=0.28\textwidth]{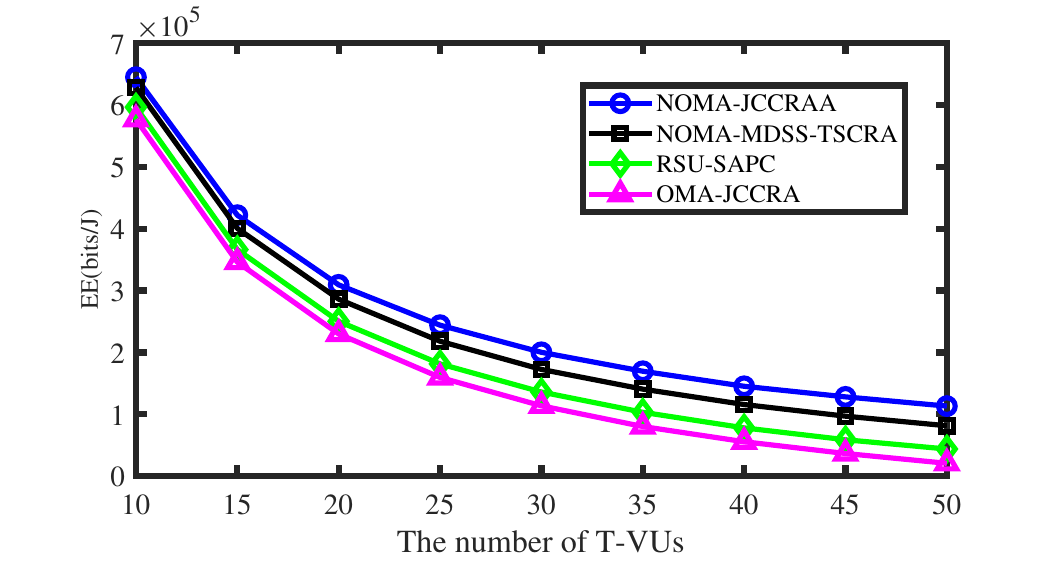}
	\caption{The system EE performance for different number of T-VUs.}
	\label{Sim1}
\end{figure}

\begin{figure}[t]
	\centering
	\includegraphics[width=0.45\textwidth,height=0.28\textwidth]{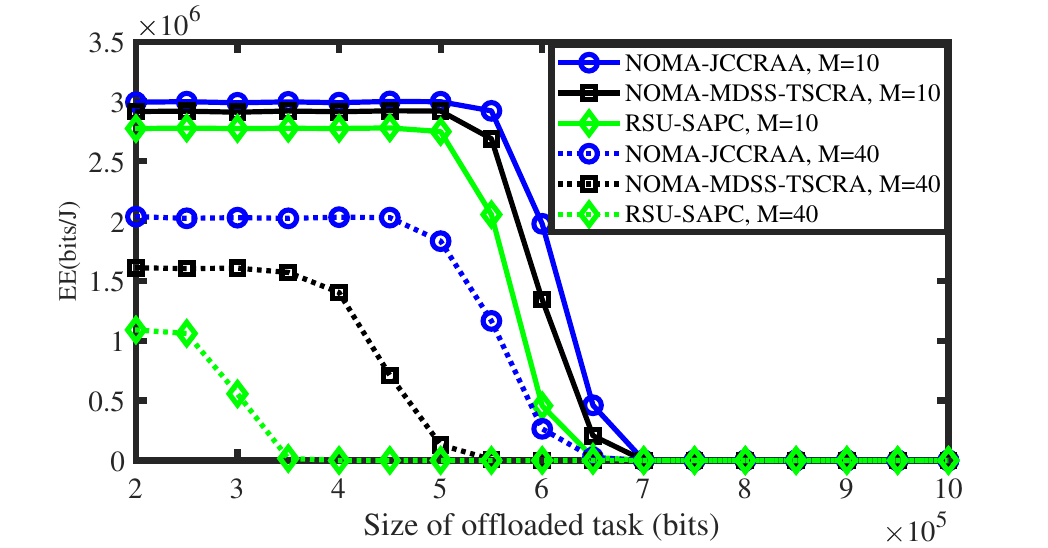}
	\caption{The system EE performance for different offloaded task sizes.}
	\label{Sim2}
\end{figure}

\begin{figure}[t]
	\centering
	\includegraphics[width=0.45\textwidth,height=0.28\textwidth]{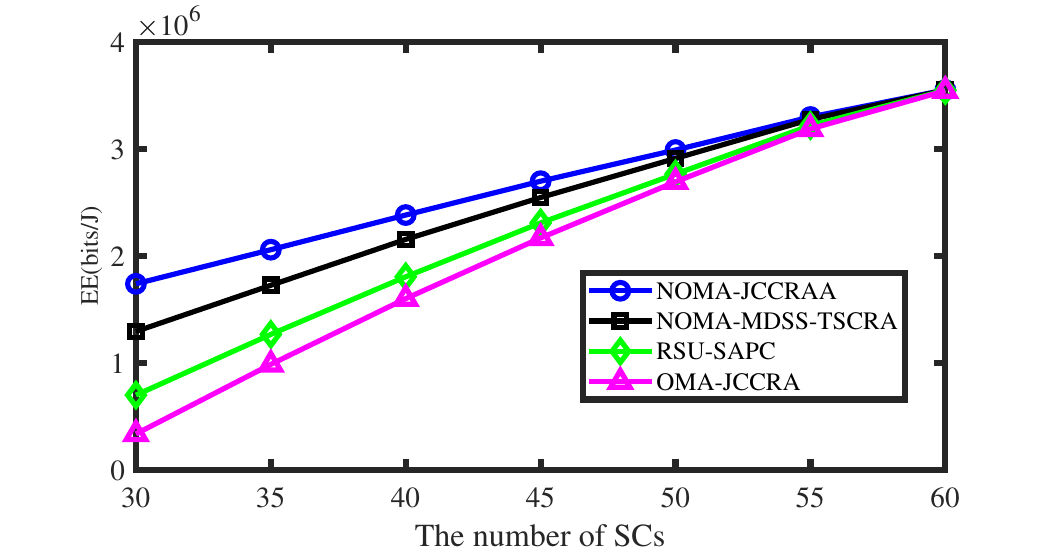}
	\caption{The system EE performance for different number of SCs.}
	\label{Sim3}
\end{figure}

The impact of the number of T-VUs, the size of the offloaded tasks and the number of SCs on the system EE are simulated, respectively. The obtained resluts are shown in Figs. \ref{Sim1}-\ref{Sim3}. To show the superiority of the proposed JCCRAA, three baselines are simulated and compared: 
	1) NOMA-MDSS-TSCRA algorithm, which is composed of the MDSS and the proposed task splitting and communication resource allocation algorithm.   
	2) RSU-SAPC algorithm, which is composed of the RSU-based offloading strategy and the proposed communication resource allocation algorithm. 
	3) OMA-JCCRA algorithm, which is adopted by the proposed JCCRA algorithm based on the orthogonal multiple access. 

The system EE for different number T-VUs is shown in \figref{Sim1}, from which we see that the EE decreases as the number of T-VUs increases for all the simulated algorithms. The reason is that, as the number of T-VUs increases, the competition for limited communication resources intensifies, resulting in severe co-channel interference and a degradation in EE performance. In addition, we see that the proposed NOMA-JCCRAA performs best, and with the social-mobility-aware algorithm, a gain of approximately 17\%-32\% can be achieved. \figref{Sim2} shows the effect of the size of offloaded tasks at each T-VU on the EE performance when the size of the offloaded tasks at each T-VU varies. The simulation results reveal that as the size of the offloaded tasks increases, the EE decreases. This is attributed to an increase in task delay, making it difficult to satisfy the constraint of delay and ultimately reducing the EE performance of the VEC network. From \figref{Sim3}, we see that the EE increases when the number of the available SCs increases from 30 to 60 for all the simulated algorithms. This is because when the number of the available SCs increases, more users can occupy the spectrum bands individually, improving the system EE. 


\section{Conclusions}
This letter focuses investigation on the social-mobility-aware EE maximization in VEC networks, where the T-VUs can offload the computation tasks to the S-VUs and the RSUs by NOMA. An EE maximization problem was formulated to jointly assign the 2C resources. Since the optimization turned out to be NP-hard, to solve it, an iterative JCCRAA was proposed. Simulation results have shown that the proposed JCCRAA not only can help appropriately allocate the communication and computation resources, but also can achieve a system EE gain of approximately 17\%-32\% by using the proposed social-mobility-aware strategy.

\appendices

\bibliographystyle{IEEEtran}

\bibliography{IEEEabrv,20230601-XueTong-NOMA_VEC}
\end{document}